\documentclass[fleqn,usenatbib]{mnras}

\usepackage{amsmath}	
\usepackage{amssymb}	
\usepackage{txfonts} 

\usepackage[T1]{fontenc}

\DeclareRobustCommand{\VAN}[3]{#2}
\let\VANthebibliography\thebibliography
\def\thebibliography{\DeclareRobustCommand{\VAN}[3]{##3}\VANthebibliography}

\usepackage{ulem}

\usepackage[dvipdfmx]{graphicx}	

\title[Stellar explosions fostered by turbulent $\alpha$-effect]{Neutrino-driven massive stellar explosions in 3D fostered by magnetic fields via turbulent $\alpha$-effect}

\author[J. Matsumoto, T. Takiwaki and K. Kotake]{Jin Matsumoto$^{1}$\thanks{Email:jin@rk.phys.keio.ac.jp, jin@kusastro.kyoto-u.ac.jp}, Tomoya Takiwaki$^{2}$ and Kei Kotake$^{3}$\\
$^{1}$Keio Institute of Pure and Applied Sciences, Keio University, Yokohama 223-8522, Japan\\
$^{2}$National Astronomical Observatory of Japan, Tokyo 181-8588, Japan\\
$^{3}$Department of Applied Physics and Research Institute of Stellar Explosive Phenomena, Fukuoka University, Fukuoka 814-0180, Japan}

\date{Accepted XXX. Received YYY; in original form ZZZ}

\pubyear{2023}

\begin{document}
\label{firstpage}
\pagerange{\pageref{firstpage}--\pageref{lastpage}}
\maketitle

\begin{abstract}
We investigate the influence of magnetic field amplification on the core-collapse supernovae in highly magnetized progenitors through three-dimensional simulations. By considering rotating models, we observe a strong correlation between the exponential growth of the magnetic field in the gain region and the initiation of shock revival, with a faster onset compared to the non-rotating model. We highlight that the mean magnetic field experiences exponential amplification as a result of $\alpha$-effect in the dynamo process, which works efficiently with the increasing kinetic helicity of the turbulence within the gain region. Our findings indicate that the significant amplification of the mean magnetic fields leads to the development of locally intense turbulent magnetic fields, particularly in the vicinity of the poles, thereby promoting the revival of the shock by neutrino heating.
\end{abstract}

\begin{keywords}
stars: massive -- stars: magnetic field -- supernovae: general
\end{keywords}


\section{Introduction}
The impact of the magnetic field on the explosion mechanisms of core-collapse supernovae (CCSNe) is a matter of significant importance.  The combination of the rapid rotation and magnetic field of a massive star leads to the launching of magnetically-driven jets during the collapse of the massive star \citep[e.g.,][]{Takiwaki09, Sawai16, Kuroda20, Obergaulinger20, Obergaulinger21, Bugli21, Bugli23,Powell23}. Even in the context of a slowly-rotating progenitor, the amplification of the magnetic field due to the turbulence is expected to foster the onset of neutrino-driven explosions \citep{Obergaulinger14, Muller20, Matsumoto22, Varma23}. From the seminal work by \cite{ewald}, one of the grand challenges in the CCSN theory has long been to unveil the role of the magnetic fields in the explosion mechanism (e.g., \citealt{Burrows21, Janka12, kk06} for review).
 
In the dynamo theory, a mean kinetic helicity of the turbulence is one possible origin to amplify the mean magnetic field drastically \citep{Brandenburg05}. This is the so-called $\alpha$-effect. The mean kinetic helicity is naturally generated in a rotating convection system by a swirling motion of the fluid through the interaction between the Coriolis force and the convection flow \citep{Spruit90, Miesch05, Miesch09}. In the context of CCSNe, the proto-neutron star (PNS) and neutrino-driven convection develop while the massive star collapses \citep[e.g.,][]{Janka12, Radice18}. Although the generation mechanism of the strong magnetic field of PNSs is still under debate \citep{Akiyama03, Guilet15a, Reboul20, Reboul22, Barrere22}, the convection dynamo is considered as a key process for the origin of the magnetic field of PNSs and has been extensively studied in the PNS convection \citep{Thompson93, Bonanno03, Rheinhardt05, Raynaud20, Masada22, White22}. In fact, recent 3D dynamo simulations report that the turbulent $\alpha$-effect, the origin of which is the kinetic helicity in the turbulent electromotive force based on the mean-field theory works for the amplification mechanism of the magnetic field in the PNS convection \citep{Raynaud20, Masada22}.

While the turbulent $\alpha$-effect has been investigated in the PNS convection, its effect based on the generation of the kinetic helicity in the neutrino-driven convection on the explosion mechanism of the rotating massive star is not fully understood yet (see also \citealt{Endeve12} for the effect of the kinetic helicity in non-rotating progenitor). In this work, we focus on the impact of the magnetic field amplification due to the turbulent $\alpha$-effect in the gain region on the explosion mechanism of CCSN in the context of the slowly rotating model through 3D magnetohydrodynamic (MHD) simulations.

This $\it Letter$ is organized as follows. In Section 2, we describe our model in our calculations. We present the main results in Section 3. Section 4 is devoted to conclusions and discussions.

\begin{figure*}
\begin{center}
\scalebox{0.85}{{\includegraphics{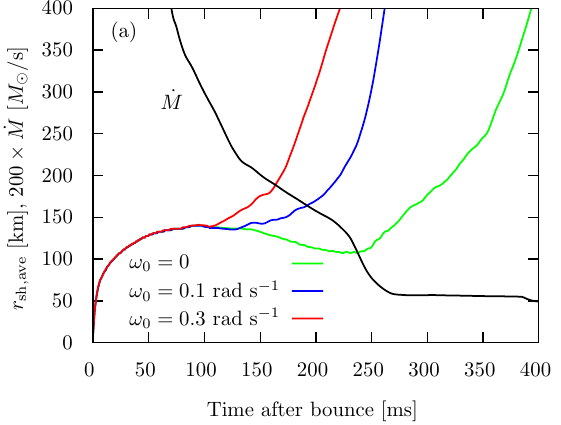}}}
\scalebox{0.85}{{\includegraphics{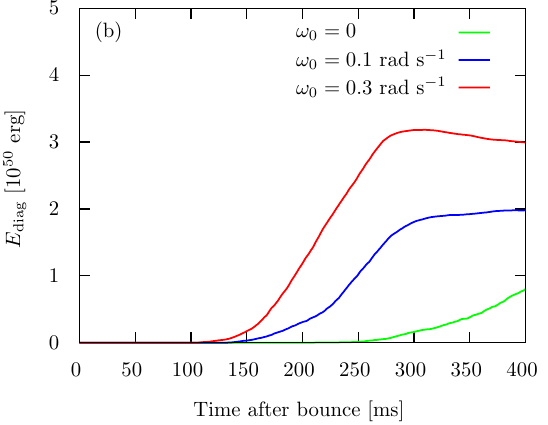}}}
\caption{Panel (a): Evolution of the averaged shock radii, $r_{\rm sh, ave}$, for all models and the mass accretion rate at $r=500$\,km, $\dot{M}$, (black line) for the non-rotating model. Panel (b): Comparison of the diagnostic explosion energy including magnetic energy defined in \citet{Matsumoto22}.}
\label{RshEexp}
\end{center}
\end{figure*}

\section{Numerical models} \label{numerical methods}
All numerical settings in this study are the same as our previous work \citep{Matsumoto22} except the initial angular velocity of the progenitor, $\omega_{\rm ini}$. The employed pre-supernova progenitor model is $27$ $M_{\odot}$ of \cite{Woosley02}, s27. We consider simple rotation profiles of s27 in a parametric manner because of the limited understanding of the spatial distribution of the rotation in the pre-collapse phase of massive stars. Following \citet{Takiwaki04} and \citet{Takiwaki09}, we assume a cylindrical rotation as follows:
\begin{eqnarray}
\omega_{\rm ini}(x,z) = \omega_0 \frac{x_0^2}{x^2 + x_0^2} \frac{z_0^4}{z^4 + z_0^4} \;, \label{eq: rotation profile}
\end{eqnarray}
where $x$ and $z$ indicate distance from the rotational axis and the equatorial plane and $x_0=z_0=1000$\,km. We set $\omega_0=0.3$, $0.1$ or $0$\, rad s$^{-1}$. The last one corresponds to a non-rotating model calculated and labelled as s27.0B12PPM5 in the previous work. The initial strength of the magnetic field in the $z$-direction of a central core of s27 is roughly $10^{12}$\,G. \cite{Matsumoto22} provides the detailed structure of the field. The equation of state of \citet{Lattimer91} with a nuclear incomprehensibility of $K = 220$\,MeV is employed. Three models that are identified by the value of $\omega_0$ are calculated using our MHD supernova code \citep[3DnSNe;][]{Takiwaki16, Matsumoto20, Matsumoto22} with $5$th-order spatial accuracy in a spherical coordinate system ($r, \theta, \phi$). The calculation domain covers a sphere whose radius is $5000$\,km with a resolution of $n_r \times n_{\theta} \times n_{\phi}$ = $480 \times 64 \times 128$. The grid-cell size to the radial direction logarithmically stretches. The resolution of the polar angle is given by $\Delta({\rm cos}\theta)=$ const. covering $0 \le \theta \le \pi$. The azimuthal angle is uniformly divided into $\Delta \phi = \pi/ 64$ covering $0 \le \phi \le 2\pi$.

\section{Results} \label{results}
Fig.~\ref{RshEexp} shows the temporal evolution of averaged shock radii (panel a) and the diagnostic explosion energy (panel b). Red, blue and green lines correspond to the models $\omega_0=0.3$, $0.1$ and $0$ rad s$^{-1}$, respectively in each panel. The evolution of the mass accretion rate for the non-rotating model evaluated at $r=500$\,km from the center of the calculation domain, $\dot{M}$, is plotted as a reference in panel (a). Note that it is not so different between all models before the shock revival because the centrifugal force in rotating models does not prevent the gravitational collapse of the s27 progenitor in the range of $\omega_0$ we assume in this work. Therefore, the core bounce time ($\sim$ 200 ms after the start of the simulation) is the same in all models.

\begin{figure*}
\begin{center}
\scalebox{0.25}{{\includegraphics{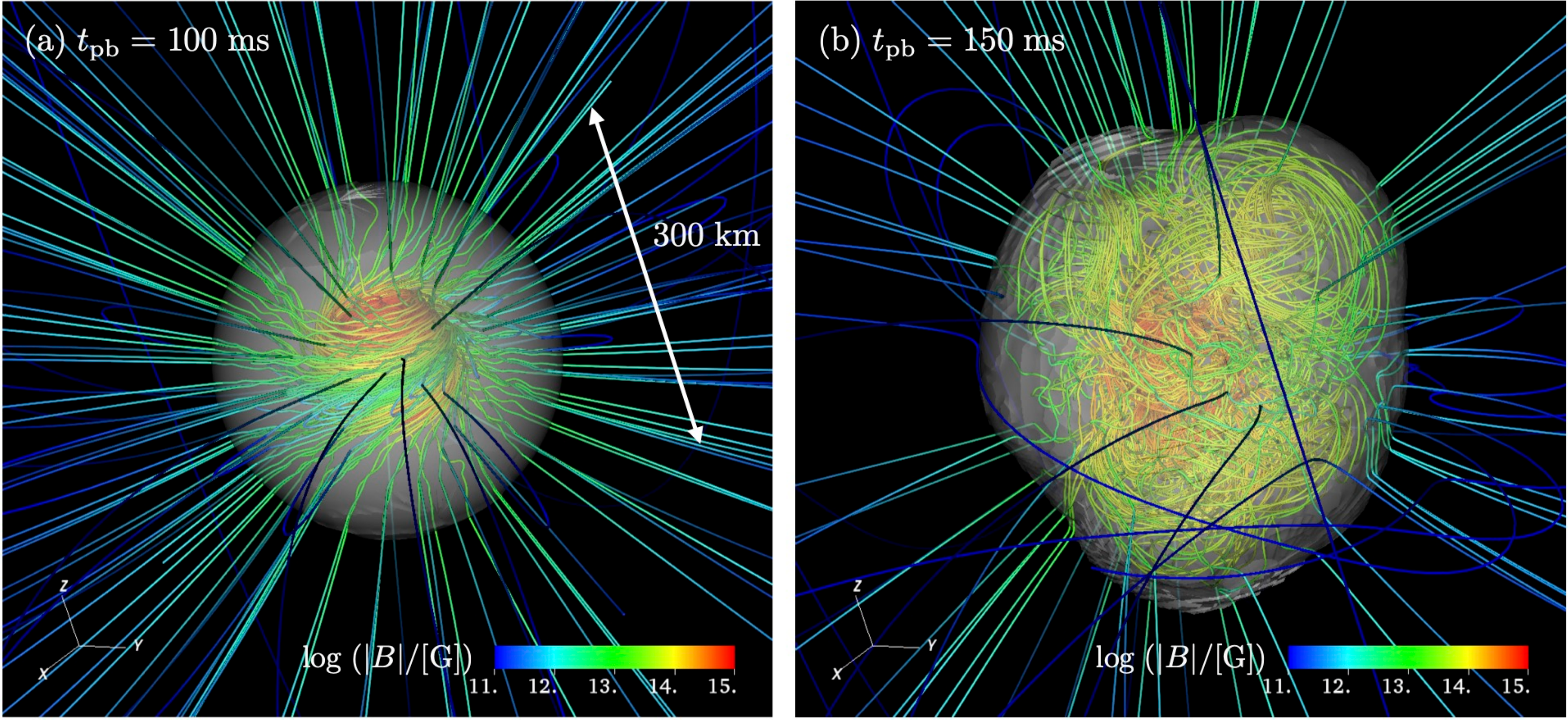}}}
\caption{Snapshots of the shock surface (whitish sphere) and magnetic field lines for model $\omega_0=0.3$ rad s$^{-1}$. Panels (a) and (b) correspond to $t_{\rm pb}=100$ and $150$\,ms, respectively. The spatial scale is represented by a white two-headed arrow in panel (a) that is parallel to the $z$-axis. The spatial scale and the viewing angle of the central object are fixed for both panels.}
\label{3Dfigure}
\end{center}
\end{figure*}

As discussed in \cite{Matsumoto22}, in the non-rotating model (green line), the shock revival occurs at the sudden drop of $\dot{M}$. The increased gas pressure behind the shock due to the neutrino heating overcomes the ram pressure of the accreting fluid onto the shock at this time and starts to push the shock outwardly because the ram pressure decreases thanks to the reduction of the density of the accreting fluid. The accumulation of the magnetic field just behind the shock contributes to the increase of the total (gas and magnetic) pressure and secondary supports the shock revival. 

The shock evolution in rotating models (red and blue lines) is obviously fast compared to that in the non-rotating model (green line). This indicates that other mechanisms would be responsible for the faster explosion of the massive star through the interaction between the rotation and the magnetic field or the efficiency of the neutrino heating is enhanced. As discussed later, the locally amplified magnetic pressure due to the turbulent $\alpha$-effect whose origin is the mean kinetic helicity in rotating models contributes to the faster explosion in this work.

On the one hand, the magnetic energy is locally larger than the thermal energy near the poles in the gain region to drive the fast explosion, but on the other hand, the total thermal energy in the gain region is large compared to the magnetic energy. Therefore, the thermal energy is dominant in the diagnostic explosion energy defined in \citet{Matsumoto22} that includes the magnetic energy. In Fig.~\ref{RshEexp}(b), the temporal evolution of the diagnostic explosion energy for all models is plotted. The diagnostic explosion energy in the faster explosion model is larger. It is reasonable because as the mass accretion rate is larger at the earlier phase of the evolution (see the black line in Fig.~\ref{RshEexp}a), the neutrino luminosity also becomes larger in the faster explosion model at the timing of the onset of the shock revival. This gives plenty of thermal energy to the expanding ejecta through neutrino heating.

When we consider the overburden energy of unshocked region in the whole s27 progenitor beyond the calculation domain \citep{Bruenn13, Bollig21, Mori23}, it is $\sim 10^{51}$\,erg at the final time of calculation ($t_{\rm pb}=400$\,ms) in all models. Therefore, the total energy (overburden and diagnostic explosion energy) of our models is negative. However, as pointed out in \cite{Marek09}, the energy release by nuclear burning in shock-heated material is expected to compensate for the overburden energy.

Fig.~\ref{3Dfigure} shows the snapshots of the shock surface and the magnetic field lines for model $\omega_0=0.3$ rad s$^{-1}$ at $t_{\rm pb}=100$\,ms (panel a) and $t_{\rm pb}=150$\,ms (panel b). Note that $t_{\rm pb}$ means post-bounce time in the calculation for CCSN. The spatial scale size (illustrated by a white two-headed arrow in panel a) and the viewing position are the same in both panels. The shock position is represented by a whitish and transparent sphere in each panel. The color of the magnetic field lines represents the absolute strength of the magnetic field. Since we initially assume the strong magnetic field ($B_{0}=10^{12}$\,G) within $r \le 1000$\,km in this work, the magnetic field near the central region of the calculation domain is simply amplified to the $B=10^{15}$\,G level after the gravitational collapse of the progenitor through the magnetic flux conservation as follows:
\begin{eqnarray}
B_{\rm PNS} \sim 10^{15}\,{\rm G} \; \Biggl ( \frac{B_0}{10^{12}\,\rm{G}} \Biggr )
\Biggl ( \frac{30\,{\rm km}}{r_{\rm PNS}} \Biggr )^2 \;. \label{eq: estimated B_PNS}
\end{eqnarray}

The neutrino-driven convection starts to develop due to the negative gradient of the entropy around at $t_{\rm pb}=100$\,ms. The evolution of the magnetic field is different before/after this phase. The gravitational collapse of the matter in the radial direction leads to the split-monopole-like configuration of the magnetic field lines except the central part of the star before the onset of the neutrino-driven convection. Although the central core of the progenitor initially rotates rigidly, the accretion of the fluid results in a strong differential rotation near the rotational axis. It contributes to the magnetic field winding around the pole in the spherical coordinate system and the generation of the toroidal component of the magnetic field. These structures of the magnetic field are confirmed in Fig.~\ref{3Dfigure}(a). After the onset of the neutrino-driven convection, it induces the non-radial motion of the fluid and the turbulence in the gain region. They interact with the magnetic field lines through the swirling motion of the fluid. In Fig.~\ref{3Dfigure}(a), slightly bent magnetic field lines are observed behind the shock (gain region). 

Fig.~\ref{3Dfigure}(b) shows the magnetic field configuration after the shock revival. Although the azimuthal component of the magnetic field is dominant near the rotational axis even in this phase, each component of the magnetic field in the gain region (behind the shock) is comparable due to the interaction of the magnetic field with the turbulence.

\begin{figure*}
\begin{center}
\scalebox{0.26}{{\includegraphics{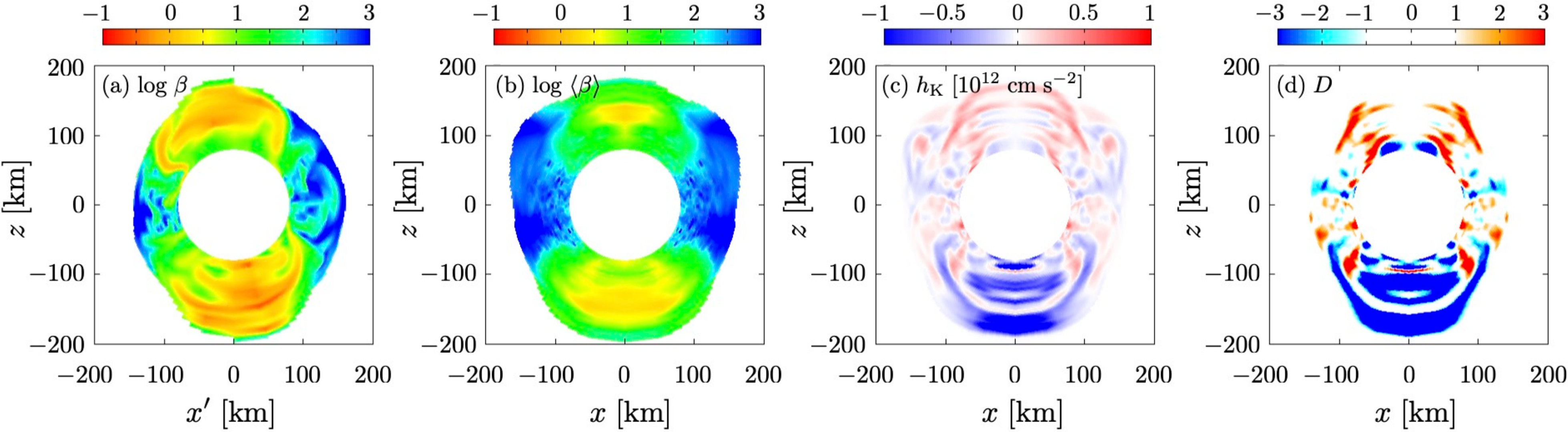}}}
\caption{2D distribution of (a) plasma $\beta$, longitudinal averages of (b) plasma $\beta$, (c) kinetic helicity and (d) dynamo number around the gain region for model $\omega_0=0.3$ at $t_{\rm pb}=139$\,ms. Note that $x^{\prime}$ in panel (a) means one axis, whose direction is along $\phi=0.025$ on $x$-$y$ plane and origin matches that of $x$. Data in the negative region of $x$ are copied from data in the positive region in panels (b), (c) and (d).}
\label{fig3}
\end{center}
\end{figure*}

Fig.~\ref{fig3} shows the 2D distribution of (a) plasma $\beta = P_{\rm gas}/P_{\rm mag}$, longitudinal averages of (b) plasma $\beta$, (c) kinetic helicity and (d) dynamo number around the gain region for model $\omega_0=0.3$\,rad s$^{-1}$ at $t_{\rm pb}=139$\,ms (after the shock revival) where $P_{\rm gas}$ and $P_{\rm mag}$ are gas and magnetic pressure, respectively. Here, the longitudinal average of variable $X$ is given by
\begin{eqnarray}
\langle X \rangle (r,\theta) = \frac{1}{2 \pi}\int_0^{2 \pi}X(r,\theta,\phi) {\rm d} \phi \;. \label{mean value}
\end{eqnarray}
In Fig.~\ref{fig3}(a), we can find the low $\beta$ region (red) near the pole where the magnetic field is fully amplified to $\mathcal{O}(10^{14})$\,G as shown in Fig.~\ref{3Dfigure}(b). On the other hand, the strength of the magnetic field at the equatorial region is $\mathcal{O}(10^{13})$\,G. In the low $\beta$ region, the magnetic pressure is larger than the gas pressure. This indicates that the magnetic pressure is the main driver for the explosion mechanism in our rotating model. However, as shown in Fig.~\ref{fig3}(b), the longitudinal average of plasma $\beta$ in the gain region is larger than unity in the meridional plane ($r,\theta$). Therefore, the volume-averaged magnetic pressure in the gain region is less than the volume-averaged gas pressure. Locally amplified magnetic pressure is expected to contribute to the outward shock expansion.

As mentioned in the introduction, the rotating convection system is expected to generate kinetic helicity and result in the amplification of the magnetic field by the dynamo action called $\alpha$-effect. To identify the amplification mechanism of the magnetic field in our rotating model, we follow the analysis of the mean-field theory of the magnetic field \citep{Brandenburg05}. The velocity and magnetic field are decomposed by the mean and turbulent components as follows:
\begin{eqnarray}
\mbox{\boldmath $v$}(r,\theta,\phi) = \langle \mbox{\boldmath $v$} \rangle (r,\theta) + \mbox{\boldmath $v$}^{\prime}(r,\theta,\phi) \; ,\\
\mbox{\boldmath $B$}(r,\theta,\phi) = \langle \mbox{\boldmath $B$} \rangle (r,\theta) + \mbox{\boldmath $B$}^{\prime}(r,\theta,\phi) \; .
\end{eqnarray}
Here, the prime represents the turbulent component. Note that we consider only the longitudinal average of the variable for the mean component and the time average is neglected for simplicity in our analysis. The mean kinetic helicity is defined by
\begin{eqnarray}
h_{\rm K}(r,\theta) = \langle \mbox{\boldmath $v$}^{\prime} \cdot \mbox{\boldmath $\omega$}^{\prime} \rangle \;,
\end{eqnarray}
where $\mbox{\boldmath $\omega$}^{\prime}$ stands for the turbulent vorticity vector given by
\begin{eqnarray}
\mbox{\boldmath $\omega$}^{\prime}(r,\theta,\phi) = \nabla \times \mbox{\boldmath $v$}^{\prime}(r,\theta,\phi) \;.
\end{eqnarray}
Under the first-order smoothing approximation,
the induction equations of the mean and turbulent magnetic field are given by
\begin{eqnarray}
\frac{\partial \langle \mbox{\boldmath $B$} \rangle}{\partial t} = \nabla \times (\langle \mbox{\boldmath $v$} \rangle \times \langle \mbox{\boldmath $B$} \rangle - \eta \nabla \times \langle \mbox{\boldmath $B$} \rangle + \mbox{\boldmath $\epsilon$}) \; , \label{eq: induction eq for mean b-field}
\end{eqnarray}
\begin{eqnarray}
\frac{\partial \mbox{\boldmath $B$}^{\prime}}{\partial t} = \nabla \times (\mbox{\boldmath $v$}^{\prime} \times \langle \mbox{\boldmath $B$} \rangle ) \; , \label{eq: induction eq for turbulent b-field}
\end{eqnarray}
respectively. Here $\mbox{\boldmath $\epsilon$} \equiv \alpha \langle \mbox{\boldmath $B$} \rangle - \eta_t \nabla \times \langle \mbox{\boldmath $B$} \rangle$ is a turbulent electromotive force and 
\begin{eqnarray}
\alpha \equiv -\frac{1}{3} \tau_{\rm cor} h_{\rm K} \; ,
\end{eqnarray}
\begin{eqnarray}
\eta_t \equiv  \frac{1}{3} \tau_{\rm cor} \langle {v^{\prime}}^2 \rangle \; ,
\end{eqnarray}
where $\tau_{\rm cor}$ is the correlation time. When the turbulent process is dominant in the amplification of the mean magnetic field, the first term of the RHS in the equation (\ref{eq: induction eq for mean b-field}) is neglected. In addition, as we set $\eta=0$ in our simulations, the induction equation for the mean magnetic field (\ref{eq: induction eq for mean b-field}) is reduced as follows:
\begin{eqnarray}
\frac{\partial \langle \mbox{\boldmath $B$} \rangle}{\partial t} = \nabla \times (\alpha \langle \mbox{\boldmath $B$} \rangle) + \eta_t \Delta \langle \mbox{\boldmath $B$} \rangle \; . \label{eq: reduce induction equation}
\end{eqnarray}
By inserting the perturbation of the form $\delta \langle \mbox{\boldmath $B$} \rangle \propto {\rm exp}(i \mbox{\boldmath $k$} \cdot \mbox{\boldmath $x$} + \sigma t)$ into the equation (\ref{eq: reduce induction equation}), we obtain the following dispersion relation:
\begin{eqnarray}
\sigma = \alpha k - \eta_t k^2 = \eta_t k^2 (D-1) \; , \label{eq: dispersion relation}
\end{eqnarray}
where $\sigma$ and $k=|\mbox{\boldmath $k$}|$ are a growth rate of the mean magnetic field and a wave number, respectively.
The condition for the exponential growth of the mean magnetic field due to the $\alpha$-effect ($\sigma > 0$) is obtained as follows:
\begin{eqnarray}
D \equiv \frac{\alpha}{k \eta_t} > 1 \;.
\end{eqnarray}
Here $D$ is a dynamo number.

From Fig.~\ref{fig3}, we can find a good correlation between the low $\beta$ region (red) and the distribution of the high absolute value of the kinetic helicity. This indicates that the kinetic helicity amplifies the magnetic field because the strength of the magnetic field in the low $\beta$ region is relatively strong compared to that in the high $\beta$ region (green and blue). In addition, we can also observe a good correlation between the 2D distribution of the kinetic helicity and dynamo number. In Fig.~\ref{fig3}(d), the coloured region represents a high dynamo number ($D > 1$) only in the gain region. Here we set $1/k=100$\,km that roughly corresponds to the length scale of the gain region, $L_{\rm gain}$. This implies that the $\alpha$-effect is responsible for the amplification of the magnetic field in our rotating model.

\begin{figure}
\begin{center}
\scalebox{0.85}{{\includegraphics{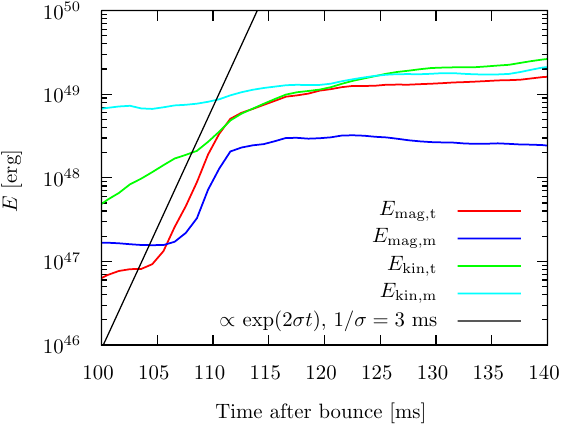}}}
\caption{Temporal evolution of the kinetic and magnetic turbulent and mean energy in the gain region around the onset of the shock revival for model $\omega_0=0.3$\,rad s$^{-1}$.}
\label{growth_rate}
\end{center}
\end{figure}

To verify the origin of the magnetic field amplification in our rotating model quantitatively, the time evolution of the mean and turbulent kinetic and magnetic energy in the gain region around the onset of the shock revival for model $\omega_0=0.3$\,rad s$^{-1}$ is shown in Fig.~\ref{growth_rate}. The mean and turbulent magnetic energy is exponentially amplified during the period between $t_{\rm }=105$ and $110$\,ms. It is closely linked to the onset of the shock revival (see red line in Fig.~\ref{RshEexp}a). Assuming a large dynamo number, the growth rate of the magnetic amplification due to the $\alpha$-effect is roughly estimated from the dispersion relation (\ref{eq: dispersion relation}) as follows:
\begin{eqnarray}
\sigma \sim \frac{\alpha}{L_{\rm gain}} \sim \frac{\tau_{\rm cor}h_{\rm K}}{3 L_{\rm gain}} \sim \frac{h_{\rm K}}{3 v_{\rm adv}} \sim \frac{1}{3} \;{\rm ms}^{-1}\;,
\end{eqnarray}
where $\tau_{\rm cor} \sim \tau_{\rm adv} \equiv L_{\rm gain} / v_{\rm adv}$. Here $\tau_{\rm adv}$ and $v_{\rm adv} \sim 10^9$ cm s$^{-1}$ are advection timescale and advection velocity in the gain region, respectively. In our rotating model, the kinetic helicity suddenly increases to the order of $10^{12}$\;cm s$^{-2}$ after around $t_{\rm pb} = 100$\;ms (see Fig.~\ref{fig3}c). The anticipated growth rate of the mean magnetic energy from the mean-field theory, $2\sigma$, is also plotted by a black line in Fig.~\ref{growth_rate}. In addition to the growth rate of the mean magnetic energy, the growth rate of the turbulent magnetic energy is roughly equal to $2\sigma$. Since, in the first-order smoothing approximation of the mean-field theory, the time evolution of the turbulent magnetic field is given by the induction equation (\ref{eq: induction eq for turbulent b-field}), it is reasonable that the growth rate of it is almost same as the growth rate of the mean magnetic field.
In this work, we consider the slowly rotating model of the progenitor and the rotational energy of the fluid is smaller than the turbulent kinetic energy in the gain region during the exponential amplification phase of the magnetic field. This implies that the convective dynamo effectively works in the magnetic amplification compared to magnetorotational instability \citep{Balbus97}. It is clear from Fig.~\ref{growth_rate} that the turbulent magnetic energy (red line) in the gain region is large compared to the mean magnetic energy (blue line). Therefore, the turbulent magnetic pressure that is amplified via $\alpha$-dynamo action of the mean magnetic field is responsible for the fast explosion in our rotating model.

\section{Summary and Discussion} \label{summary}
The impact of magnetic field amplification on the explosion mechanism of a strongly magnetized $27$ $M_{\odot}$ pre-supernova progenitor is studied by performing 3D MHD supernova simulations in the context of the slowly-rotating progenitor models. We find that the exponential growth of the magnetic field in the gain region in our rotating models has a good correlation with the onset of the shock revival, which is faster than that of the non-rotating model. In addition, the explosion energy in the more rapidly rotating model is larger. We point out that the mean magnetic field is exponentially amplified due to the $\alpha$-effect in the dynamo process, which works efficiently with the growing kinetic helicity of the turbulence in the rotating gain region. Our results show that the drastic amplification of the mean magnetic fields leads to the generation of localized and strong turbulent magnetic fields near poles in the gain region, promoting the shock revival by neutrino heating.

Obviously, we need more sophisticated numerical methods to obtain more accurate predictions to model shock revival through neutrino-heating in multi-dimension, as has been established by various studies \citep[e.g.,][]{Buras06, Suwa10, Takiwaki12, Nakamura15, Pan16, Summa16, OConnor18a, Burrows20, Nagakura20, Bollig21, Vartanyan22}. Despite the existence of some general relativistic simulations \citep{Mueller12, OConnor18b, Kuroda22, Rahman22}, the majority of the studies use phenomenological GR potential. The ab-initio neutrino transport requires huge computational resources \citep{iwakami20} and approximate methods are usually employed. Furthermore, the effect of neutrino oscillation should be taken into account \citep{Nagakura23, Ehring23} and the impact of neutrino reactions remains to be examined \citep{Kotake18, Sugiura2022}. Moreover, properties of high density equation of state are still under debate \citep{Fischer14, Fischer20}.

In this {\it Letter}, we have proposed the so-far-unidentified mechanism to forge the explosion onset via the turbulent $\alpha$-effect for a strongly magnetized and slowly-rotating progenitor. The diversity of supernova explosions \citep{Kulkarni12,Taddia18, Martinez22b} might originate from the property of the progenitors \citep[e.g.,][]{Smartt09, Aguilera-Dena20, Takahashi21}. Especially, the binary evolution of stars is important because it may give complicated features and a wide range of spin parameters for the progenitor, which are not established in a single star evolution through mass transfer from a companion, tidal locking between binaries or stellar mergers \citep{Cantiello07, Chatzopoulos20}. Recent advancements in 3D MHD simulations of the stellar evolution enable us to unveil the dynamics and evolution of the system where the rotation of the progenitor interacts with the magnetic field \citep{Varma21, Yoshida21, Mcneill22, Fields22}. In order to comprehend the origin of the diversity of CCSNe, sophisticated and a wide range of parameter studies for the magnetic field and spin period of the progenitor are needed, though computationally expensive, towards which we, including CCSN modellers in the globe are making the steady step.

\section*{Acknowledgements}
We thank Y. Masada, K. Nakamura, Y. Asahina, Y. Suwa and A. Harada for informative and stimulating discussions.
Numerical computations were carried out on Cray XC50 at the Center for Computational Astrophysics, National Astronomical Observatory of Japan and on Cray XC40 at YITP in Kyoto University. This work was supported by the Keio Institute of Pure and Applied Sciences (KiPAS) project at Keio University, Research Institute of Stellar Explosive Phenomena at Fukuoka University \& also from the University grant No.GR2302,
and also by JSPS KAKENHI
Grant Number (JP19K23443, 
JP20K14473, 
JP21H01088, 
JP22H01223, 
JP23H01199, 
and JP23K03400). 
This research was also supported by MEXT as “Program for Promoting researches on the Supercomputer Fugaku” (Structure and Evolution of the Universe Unraveled by Fusion of Simulation and AI; Grant Number JPMXP1020230406) and JICFuS.

\section*{DATA AVAILABILITY}
The data underlying this article will be shared on reasonable request to the corresponding author.

\bibliographystyle{mnras}
\bibliography{papers} 

\begin{thebibliography}{}
\makeatletter
\relax
\def\mn@urlcharsother{\let\do\@makeother \do\$\do\&\do\#\do\^\do\_\do\%\do\~}
\def\mn@doi{\begingroup\mn@urlcharsother \@ifnextchar [ {\mn@doi@}
  {\mn@doi@[]}}
\def\mn@doi@[#1]#2{\def\@tempa{#1}\ifx\@tempa\@empty \href
  {http://dx.doi.org/#2} {doi:#2}\else \href {http://dx.doi.org/#2} {#1}\fi
  \endgroup}
\def\mn@eprint#1#2{\mn@eprint@#1:#2::\@nil}
\def\mn@eprint@arXiv#1{\href {http://arxiv.org/abs/#1} {{\tt arXiv:#1}}}
\def\mn@eprint@dblp#1{\href {http://dblp.uni-trier.de/rec/bibtex/#1.xml}
  {dblp:#1}}
\def\mn@eprint@#1:#2:#3:#4\@nil{\def\@tempa {#1}\def\@tempb {#2}\def\@tempc
  {#3}\ifx \@tempc \@empty \let \@tempc \@tempb \let \@tempb \@tempa \fi \ifx
  \@tempb \@empty \def\@tempb {arXiv}\fi \@ifundefined
  {mn@eprint@\@tempb}{\@tempb:\@tempc}{\expandafter \expandafter \csname
  mn@eprint@\@tempb\endcsname \expandafter{\@tempc}}}

\bibitem[\protect\citeauthoryear{{Aguilera-Dena}, {Langer}, {Antoniadis}  \&
  {M{\"u}ller}}{{Aguilera-Dena} et~al.}{2020}]{Aguilera-Dena20}
{Aguilera-Dena} D.~R.,  {Langer} N.,  {Antoniadis} J.,   {M{\"u}ller} B.,
  2020, \mn@doi [\apj] {10.3847/1538-4357/abb138}, \href
  {https://ui.adsabs.harvard.edu/abs/2020ApJ...901..114A} {901, 114}

\bibitem[\protect\citeauthoryear{{Akiyama}, {Wheeler}, {Meier}  \&
  {Lichtenstadt}}{{Akiyama} et~al.}{2003}]{Akiyama03}
{Akiyama} S.,  {Wheeler} J.~C.,  {Meier} D.~L.,   {Lichtenstadt} I.,  2003,
  \mn@doi [\apj] {10.1086/344135}, \href
  {https://ui.adsabs.harvard.edu/abs/2003ApJ...584..954A} {584, 954}

\bibitem[\protect\citeauthoryear{{Balbus} \& {Hawley}}{{Balbus} \&
  {Hawley}}{1998}]{Balbus97}
{Balbus} S.~A.,  {Hawley} J.~F.,  1998, \mn@doi [Reviews of Modern Physics]
  {10.1103/RevModPhys.70.1}, \href
  {https://ui.adsabs.harvard.edu/abs/1998RvMP...70....1B} {70, 1}

\bibitem[\protect\citeauthoryear{{Barr{\`e}re}, {Guilet}, {Reboul-Salze},
  {Raynaud}  \& {Janka}}{{Barr{\`e}re} et~al.}{2022}]{Barrere22}
{Barr{\`e}re} P.,  {Guilet} J.,  {Reboul-Salze} A.,  {Raynaud} R.,   {Janka}
  H.~T.,  2022, \mn@doi [\aap] {10.1051/0004-6361/202244172}, \href
  {https://ui.adsabs.harvard.edu/abs/2022A&A...668A..79B} {668, A79}

\bibitem[\protect\citeauthoryear{{Bollig}, {Yadav}, {Kresse}, {Janka},
  {M{\"u}ller}  \& {Heger}}{{Bollig} et~al.}{2021}]{Bollig21}
{Bollig} R.,  {Yadav} N.,  {Kresse} D.,  {Janka} H.-T.,  {M{\"u}ller} B.,
  {Heger} A.,  2021, \mn@doi [\apj] {10.3847/1538-4357/abf82e}, \href
  {https://ui.adsabs.harvard.edu/abs/2021ApJ...915...28B} {915, 28}

\bibitem[\protect\citeauthoryear{{Bonanno}, {Rezzolla}  \& {Urpin}}{{Bonanno}
  et~al.}{2003}]{Bonanno03}
{Bonanno} A.,  {Rezzolla} L.,   {Urpin} V.,  2003, \mn@doi [\aap]
  {10.1051/0004-6361:20031459}, \href
  {https://ui.adsabs.harvard.edu/abs/2003A&A...410L..33B} {410, L33}

\bibitem[\protect\citeauthoryear{{Brandenburg} \& {Subramanian}}{{Brandenburg}
  \& {Subramanian}}{2005}]{Brandenburg05}
{Brandenburg} A.,  {Subramanian} K.,  2005, \mn@doi [\physrep]
  {10.1016/j.physrep.2005.06.005}, \href
  {https://ui.adsabs.harvard.edu/abs/2005PhR...417....1B} {417, 1}

\bibitem[\protect\citeauthoryear{{Bruenn} et~al.,}{{Bruenn}
  et~al.}{2013}]{Bruenn13}
{Bruenn} S.~W.,  et~al., 2013, \mn@doi [\apjl] {10.1088/2041-8205/767/1/L6},
  \href {https://ui.adsabs.harvard.edu/abs/2013ApJ...767L...6B} {767, L6}

\bibitem[\protect\citeauthoryear{{Bugli}, {Guilet}  \& {Obergaulinger}}{{Bugli}
  et~al.}{2021}]{Bugli21}
{Bugli} M.,  {Guilet} J.,   {Obergaulinger} M.,  2021, \mn@doi [\mnras]
  {10.1093/mnras/stab2161}, \href
  {https://ui.adsabs.harvard.edu/abs/2021MNRAS.507..443B} {507, 443}

\bibitem[\protect\citeauthoryear{{Bugli}, {Guilet}, {Foglizzo}  \&
  {Obergaulinger}}{{Bugli} et~al.}{2023}]{Bugli23}
{Bugli} M.,  {Guilet} J.,  {Foglizzo} T.,   {Obergaulinger} M.,  2023, \mn@doi
  [\mnras] {10.1093/mnras/stad496}, \href
  {https://ui.adsabs.harvard.edu/abs/2023MNRAS.520.5622B} {520, 5622}

\bibitem[\protect\citeauthoryear{{Buras}, {Rampp}, {Janka}  \&
  {Kifonidis}}{{Buras} et~al.}{2006}]{Buras06}
{Buras} R.,  {Rampp} M.,  {Janka} H.~T.,   {Kifonidis} K.,  2006, \mn@doi
  [\aap] {10.1051/0004-6361:20053783}, \href
  {https://ui.adsabs.harvard.edu/abs/2006A&A...447.1049B} {447, 1049}

\bibitem[\protect\citeauthoryear{{Burrows} \& {Vartanyan}}{{Burrows} \&
  {Vartanyan}}{2021}]{Burrows21}
{Burrows} A.,  {Vartanyan} D.,  2021, \mn@doi [\nat]
  {10.1038/s41586-020-03059-w}, \href
  {https://ui.adsabs.harvard.edu/abs/2021Natur.589...29B} {589, 29}

\bibitem[\protect\citeauthoryear{{Burrows}, {Radice}, {Vartanyan}, {Nagakura},
  {Skinner}  \& {Dolence}}{{Burrows} et~al.}{2020}]{Burrows20}
{Burrows} A.,  {Radice} D.,  {Vartanyan} D.,  {Nagakura} H.,  {Skinner} M.~A.,
   {Dolence} J.~C.,  2020, \mn@doi [\mnras] {10.1093/mnras/stz3223}, \href
  {https://ui.adsabs.harvard.edu/abs/2020MNRAS.491.2715B} {491, 2715}

\bibitem[\protect\citeauthoryear{{Cantiello}, {Yoon}, {Langer}  \&
  {Livio}}{{Cantiello} et~al.}{2007}]{Cantiello07}
{Cantiello} M.,  {Yoon} S.~C.,  {Langer} N.,   {Livio} M.,  2007, \mn@doi
  [\aap] {10.1051/0004-6361:20077115}, \href
  {https://ui.adsabs.harvard.edu/abs/2007A&A...465L..29C} {465, L29}

\bibitem[\protect\citeauthoryear{{Chatzopoulos}, {Frank}, {Marcello}  \&
  {Clayton}}{{Chatzopoulos} et~al.}{2020}]{Chatzopoulos20}
{Chatzopoulos} E.,  {Frank} J.,  {Marcello} D.~C.,   {Clayton} G.~C.,  2020,
  \mn@doi [\apj] {10.3847/1538-4357/ab91bb}, \href
  {https://ui.adsabs.harvard.edu/abs/2020ApJ...896...50C} {896, 50}

\bibitem[\protect\citeauthoryear{{Ehring}, {Abbar}, {Janka}, {Raffelt}  \&
  {Tamborra}}{{Ehring} et~al.}{2023}]{Ehring23}
{Ehring} J.,  {Abbar} S.,  {Janka} H.-T.,  {Raffelt} G.,   {Tamborra} I.,
  2023, \mn@doi [\prd] {10.1103/PhysRevD.107.103034}, \href
  {https://ui.adsabs.harvard.edu/abs/2023PhRvD.107j3034E} {107, 103034}

\bibitem[\protect\citeauthoryear{{Endeve}, {Cardall}, {Budiardja}, {Beck},
  {Bejnood}, {Toedte}, {Mezzacappa}  \& {Blondin}}{{Endeve}
  et~al.}{2012}]{Endeve12}
{Endeve} E.,  {Cardall} C.~Y.,  {Budiardja} R.~D.,  {Beck} S.~W.,  {Bejnood}
  A.,  {Toedte} R.~J.,  {Mezzacappa} A.,   {Blondin} J.~M.,  2012, \mn@doi
  [\apj] {10.1088/0004-637X/751/1/26}, \href
  {https://ui.adsabs.harvard.edu/abs/2012ApJ...751...26E} {751, 26}

\bibitem[\protect\citeauthoryear{{Fields}}{{Fields}}{2022}]{Fields22}
{Fields} C.~E.,  2022, \mn@doi [\apjl] {10.3847/2041-8213/ac460c}, \href
  {https://ui.adsabs.harvard.edu/abs/2022ApJ...924L..15F} {924, L15}

\bibitem[\protect\citeauthoryear{{Fischer}, {Hempel}, {Sagert}, {Suwa}  \&
  {Schaffner-Bielich}}{{Fischer} et~al.}{2014}]{Fischer14}
{Fischer} T.,  {Hempel} M.,  {Sagert} I.,  {Suwa} Y.,   {Schaffner-Bielich} J.,
   2014, \mn@doi [European Physical Journal A] {10.1140/epja/i2014-14046-5},
  \href {https://ui.adsabs.harvard.edu/abs/2014EPJA...50...46F} {50, 46}

\bibitem[\protect\citeauthoryear{{Fischer}, {Wu}, {Wehmeyer}, {Bastian},
  {Mart{\'\i}nez-Pinedo}  \& {Thielemann}}{{Fischer} et~al.}{2020}]{Fischer20}
{Fischer} T.,  {Wu} M.-R.,  {Wehmeyer} B.,  {Bastian} N.-U.~F.,
  {Mart{\'\i}nez-Pinedo} G.,   {Thielemann} F.-K.,  2020, \mn@doi [\apj]
  {10.3847/1538-4357/ab86b0}, \href
  {https://ui.adsabs.harvard.edu/abs/2020ApJ...894....9F} {894, 9}

\bibitem[\protect\citeauthoryear{{Guilet}, {M{\"u}ller}  \& {Janka}}{{Guilet}
  et~al.}{2015}]{Guilet15a}
{Guilet} J.,  {M{\"u}ller} E.,   {Janka} H.-T.,  2015, \mn@doi [\mnras]
  {10.1093/mnras/stu2550}, \href
  {https://ui.adsabs.harvard.edu/abs/2015MNRAS.447.3992G} {447, 3992}

\bibitem[\protect\citeauthoryear{{Iwakami}, {Okawa}, {Nagakura}, {Harada},
  {Furusawa}, {Sumiyoshi}, {Matsufuru}  \& {Yamada}}{{Iwakami}
  et~al.}{2020}]{iwakami20}
{Iwakami} W.,  {Okawa} H.,  {Nagakura} H.,  {Harada} A.,  {Furusawa} S.,
  {Sumiyoshi} K.,  {Matsufuru} H.,   {Yamada} S.,  2020, \mn@doi [\apj]
  {10.3847/1538-4357/abb8cf}, \href
  {https://ui.adsabs.harvard.edu/abs/2020ApJ...903...82I} {903, 82}

\bibitem[\protect\citeauthoryear{{Janka}}{{Janka}}{2012}]{Janka12}
{Janka} H.-T.,  2012, \mn@doi [Annual Review of Nuclear and Particle Science]
  {10.1146/annurev-nucl-102711-094901}, \href
  {https://ui.adsabs.harvard.edu/abs/2012ARNPS..62..407J} {62, 407}

\bibitem[\protect\citeauthoryear{{Kotake}, {Sato}  \& {Takahashi}}{{Kotake}
  et~al.}{2006}]{kk06}
{Kotake} K.,  {Sato} K.,   {Takahashi} K.,  2006, \mn@doi [Reports on Progress
  in Physics] {10.1088/0034-4885/69/4/R03}, \href
  {https://ui.adsabs.harvard.edu/abs/2006RPPh...69..971K} {69, 971}

\bibitem[\protect\citeauthoryear{{Kotake}, {Takiwaki}, {Fischer}, {Nakamura}
  \& {Mart{\'\i}nez-Pinedo}}{{Kotake} et~al.}{2018}]{Kotake18}
{Kotake} K.,  {Takiwaki} T.,  {Fischer} T.,  {Nakamura} K.,
  {Mart{\'\i}nez-Pinedo} G.,  2018, \mn@doi [\apj] {10.3847/1538-4357/aaa716},
  \href {https://ui.adsabs.harvard.edu/abs/2018ApJ...853..170K} {853, 170}

\bibitem[\protect\citeauthoryear{{Kulkarni}}{{Kulkarni}}{2012}]{Kulkarni12}
{Kulkarni} S.~R.,  2012, \mn@doi [IAUS] {10.1017/S174392131200021X}, \href
  {https://ui.adsabs.harvard.edu/abs/2012IAUS..285...55K} {285, 55}

\bibitem[\protect\citeauthoryear{{Kuroda}, {Arcones}, {Takiwaki}  \&
  {Kotake}}{{Kuroda} et~al.}{2020}]{Kuroda20}
{Kuroda} T.,  {Arcones} A.,  {Takiwaki} T.,   {Kotake} K.,  2020, \mn@doi
  [\apj] {10.3847/1538-4357/ab9308}, \href
  {https://ui.adsabs.harvard.edu/abs/2020ApJ...896..102K} {896, 102}

\bibitem[\protect\citeauthoryear{{Kuroda}, {Fischer}, {Takiwaki}  \&
  {Kotake}}{{Kuroda} et~al.}{2022}]{Kuroda22}
{Kuroda} T.,  {Fischer} T.,  {Takiwaki} T.,   {Kotake} K.,  2022, \mn@doi
  [\apj] {10.3847/1538-4357/ac31a8}, \href
  {https://ui.adsabs.harvard.edu/abs/2022ApJ...924...38K} {924, 38}

\bibitem[\protect\citeauthoryear{{Lattimer} \& {Swesty}}{{Lattimer} \&
  {Swesty}}{1991}]{Lattimer91}
{Lattimer} J.~M.,  {Swesty} D.~F.,  1991, \mn@doi [\nphysa]
  {10.1016/0375-9474(91)90452-C}, \href
  {https://ui.adsabs.harvard.edu/abs/1991NuPhA.535..331L} {535, 331}

\bibitem[\protect\citeauthoryear{{Marek} \& {Janka}}{{Marek} \&
  {Janka}}{2009}]{Marek09}
{Marek} A.,  {Janka} H.~T.,  2009, \mn@doi [\apj]
  {10.1088/0004-637X/694/1/664}, \href
  {https://ui.adsabs.harvard.edu/abs/2009ApJ...694..664M} {694, 664}

\bibitem[\protect\citeauthoryear{{Martinez} et~al.,}{{Martinez}
  et~al.}{2022}]{Martinez22b}
{Martinez} L.,  et~al., 2022, \mn@doi [\aap] {10.1051/0004-6361/202142076},
  \href {https://ui.adsabs.harvard.edu/abs/2022A&A...660A..41M} {660, A41}

\bibitem[\protect\citeauthoryear{{Masada}, {Takiwaki}  \& {Kotake}}{{Masada}
  et~al.}{2022}]{Masada22}
{Masada} Y.,  {Takiwaki} T.,   {Kotake} K.,  2022, \mn@doi [\apj]
  {10.3847/1538-4357/ac34f6}, \href
  {https://ui.adsabs.harvard.edu/abs/2022ApJ...924...75M} {924, 75}

\bibitem[\protect\citeauthoryear{{Matsumoto}, {Takiwaki}, {Kotake}, {Asahina}
  \& {Takahashi}}{{Matsumoto} et~al.}{2020}]{Matsumoto20}
{Matsumoto} J.,  {Takiwaki} T.,  {Kotake} K.,  {Asahina} Y.,   {Takahashi}
  H.~R.,  2020, \mn@doi [\mnras] {10.1093/mnras/staa3095}, \href
  {https://ui.adsabs.harvard.edu/abs/2020MNRAS.499.4174M} {499, 4174}

\bibitem[\protect\citeauthoryear{{Matsumoto}, {Asahina}, {Takiwaki}, {Kotake}
  \& {Takahashi}}{{Matsumoto} et~al.}{2022}]{Matsumoto22}
{Matsumoto} J.,  {Asahina} Y.,  {Takiwaki} T.,  {Kotake} K.,   {Takahashi}
  H.~R.,  2022, \mn@doi [\mnras] {10.1093/mnras/stac2335}, \href
  {https://ui.adsabs.harvard.edu/abs/2022MNRAS.516.1752M} {516, 1752}

\bibitem[\protect\citeauthoryear{{McNeill} \& {M{\"u}ller}}{{McNeill} \&
  {M{\"u}ller}}{2022}]{Mcneill22}
{McNeill} L.~O.,  {M{\"u}ller} B.,  2022, \mn@doi [\mnras]
  {10.1093/mnras/stab3076}, \href
  {https://ui.adsabs.harvard.edu/abs/2022MNRAS.509..818M} {509, 818}

\bibitem[\protect\citeauthoryear{{Miesch}}{{Miesch}}{2005}]{Miesch05}
{Miesch} M.~S.,  2005, \mn@doi [Living Reviews in Solar Physics]
  {10.12942/lrsp-2005-1}, \href
  {https://ui.adsabs.harvard.edu/abs/2005LRSP....2....1M} {2, 1}

\bibitem[\protect\citeauthoryear{{Miesch} \& {Toomre}}{{Miesch} \&
  {Toomre}}{2009}]{Miesch09}
{Miesch} M.~S.,  {Toomre} J.,  2009, \mn@doi [Annual Review of Fluid Mechanics]
  {10.1146/annurev.fluid.010908.165215}, \href
  {https://ui.adsabs.harvard.edu/abs/2009AnRFM..41..317M} {41, 317}

\bibitem[\protect\citeauthoryear{{Mori}, {Takiwaki}, {Kotake}  \&
  {Horiuchi}}{{Mori} et~al.}{2023}]{Mori23}
{Mori} K.,  {Takiwaki} T.,  {Kotake} K.,   {Horiuchi} S.,  2023, \mn@doi [\prd]
  {10.1103/PhysRevD.108.063027}, \href
  {https://ui.adsabs.harvard.edu/abs/2023PhRvD.108f3027M} {108, 063027}

\bibitem[\protect\citeauthoryear{{M\"uller} \& {Hillebrandt}}{{M\"uller} \&
  {Hillebrandt}}{1979}]{ewald}
{M\"uller} E.,  {Hillebrandt} W.,  1979, \aap, \href
  {https://ui.adsabs.harvard.edu/abs/1979A&A....80..147M} {80, 147}

\bibitem[\protect\citeauthoryear{{M{\"u}ller} \& {Varma}}{{M{\"u}ller} \&
  {Varma}}{2020}]{Muller20}
{M{\"u}ller} B.,  {Varma} V.,  2020, arXiv e-prints, \href
  {https://ui.adsabs.harvard.edu/abs/2020arXiv200704775M} {p. arXiv:2007.04775}

\bibitem[\protect\citeauthoryear{{M{\"u}ller}, {Janka}  \&
  {Marek}}{{M{\"u}ller} et~al.}{2012}]{Mueller12}
{M{\"u}ller} B.,  {Janka} H.-T.,   {Marek} A.,  2012, \mn@doi [\apj]
  {10.1088/0004-637X/756/1/84}, \href
  {https://ui.adsabs.harvard.edu/abs/2012ApJ...756...84M} {756, 84}

\bibitem[\protect\citeauthoryear{{Nagakura}}{{Nagakura}}{2023}]{Nagakura23}
{Nagakura} H.,  2023, \mn@doi [\prl] {10.1103/PhysRevLett.130.211401}, \href
  {https://ui.adsabs.harvard.edu/abs/2023PhRvL.130u1401N} {130, 211401}

\bibitem[\protect\citeauthoryear{{Nagakura}, {Burrows}, {Radice}  \&
  {Vartanyan}}{{Nagakura} et~al.}{2020}]{Nagakura20}
{Nagakura} H.,  {Burrows} A.,  {Radice} D.,   {Vartanyan} D.,  2020, \mn@doi
  [\mnras] {10.1093/mnras/staa261}, \href
  {https://ui.adsabs.harvard.edu/abs/2020MNRAS.492.5764N} {492, 5764}

\bibitem[\protect\citeauthoryear{{Nakamura}, {Takiwaki}, {Kuroda}  \&
  {Kotake}}{{Nakamura} et~al.}{2015}]{Nakamura15}
{Nakamura} K.,  {Takiwaki} T.,  {Kuroda} T.,   {Kotake} K.,  2015, \mn@doi
  [\pasj] {10.1093/pasj/psv073}, \href
  {https://ui.adsabs.harvard.edu/abs/2015PASJ...67..107N} {67, 107}

\bibitem[\protect\citeauthoryear{{O'Connor} \& {Couch}}{{O'Connor} \&
  {Couch}}{2018a}]{OConnor18a}
{O'Connor} E.~P.,  {Couch} S.~M.,  2018a, \mn@doi [\apj]
  {10.3847/1538-4357/aaa893}, \href
  {https://ui.adsabs.harvard.edu/abs/2018ApJ...854...63O} {854, 63}

\bibitem[\protect\citeauthoryear{{O'Connor} \& {Couch}}{{O'Connor} \&
  {Couch}}{2018b}]{OConnor18b}
{O'Connor} E.~P.,  {Couch} S.~M.,  2018b, \mn@doi [\apj]
  {10.3847/1538-4357/aadcf7}, \href
  {https://ui.adsabs.harvard.edu/abs/2018ApJ...865...81O} {865, 81}

\bibitem[\protect\citeauthoryear{{Obergaulinger} \& {Aloy}}{{Obergaulinger} \&
  {Aloy}}{2020}]{Obergaulinger20}
{Obergaulinger} M.,  {Aloy} M.~{\'A}.,  2020, \mn@doi [\mnras]
  {10.1093/mnras/staa096}, \href
  {https://ui.adsabs.harvard.edu/abs/2020MNRAS.492.4613O} {492, 4613}

\bibitem[\protect\citeauthoryear{{Obergaulinger} \& {Aloy}}{{Obergaulinger} \&
  {Aloy}}{2021}]{Obergaulinger21}
{Obergaulinger} M.,  {Aloy} M.~{\'A}.,  2021, \mn@doi [\mnras]
  {10.1093/mnras/stab295}, \href
  {https://ui.adsabs.harvard.edu/abs/2021MNRAS.503.4942O} {503, 4942}

\bibitem[\protect\citeauthoryear{{Obergaulinger}, {Janka}  \&
  {Aloy}}{{Obergaulinger} et~al.}{2014}]{Obergaulinger14}
{Obergaulinger} M.,  {Janka} H.~T.,   {Aloy} M.~A.,  2014, \mn@doi [\mnras]
  {10.1093/mnras/stu1969}, \href
  {https://ui.adsabs.harvard.edu/abs/2014MNRAS.445.3169O} {445, 3169}

\bibitem[\protect\citeauthoryear{{Pan}, {Liebend{\"o}rfer}, {Hempel}  \&
  {Thielemann}}{{Pan} et~al.}{2016}]{Pan16}
{Pan} K.-C.,  {Liebend{\"o}rfer} M.,  {Hempel} M.,   {Thielemann} F.-K.,  2016,
  \mn@doi [\apj] {10.3847/0004-637X/817/1/72}, \href
  {https://ui.adsabs.harvard.edu/abs/2016ApJ...817...72P} {817, 72}

\bibitem[\protect\citeauthoryear{{Powell}, {M{\"u}ller}, {Aguilera-Dena}  \&
  {Langer}}{{Powell} et~al.}{2023}]{Powell23}
{Powell} J.,  {M{\"u}ller} B.,  {Aguilera-Dena} D.~R.,   {Langer} N.,  2023,
  \mn@doi [\mnras] {10.1093/mnras/stad1292}, \href
  {https://ui.adsabs.harvard.edu/abs/2023MNRAS.522.6070P} {522, 6070}

\bibitem[\protect\citeauthoryear{{Radice}, {Abdikamalov}, {Ott}, {M{\"o}sta},
  {Couch}  \& {Roberts}}{{Radice} et~al.}{2018}]{Radice18}
{Radice} D.,  {Abdikamalov} E.,  {Ott} C.~D.,  {M{\"o}sta} P.,  {Couch} S.~M.,
   {Roberts} L.~F.,  2018, \mn@doi [Journal of Physics G Nuclear Physics]
  {10.1088/1361-6471/aab872}, \href
  {https://ui.adsabs.harvard.edu/abs/2018JPhG...45e3003R} {45, 053003}

\bibitem[\protect\citeauthoryear{{Rahman}, {Janka}, {Stockinger}  \&
  {Woosley}}{{Rahman} et~al.}{2022}]{Rahman22}
{Rahman} N.,  {Janka} H.~T.,  {Stockinger} G.,   {Woosley} S.~E.,  2022,
  \mn@doi [\mnras] {10.1093/mnras/stac758}, \href
  {https://ui.adsabs.harvard.edu/abs/2022MNRAS.512.4503R} {512, 4503}

\bibitem[\protect\citeauthoryear{{Raynaud}, {Guilet}, {Janka}  \&
  {Gastine}}{{Raynaud} et~al.}{2020}]{Raynaud20}
{Raynaud} R.,  {Guilet} J.,  {Janka} H.-T.,   {Gastine} T.,  2020, \mn@doi
  [Science Advances] {10.1126/sciadv.aay2732}, \href
  {https://ui.adsabs.harvard.edu/abs/2020SciA....6.2732R} {6, eaay2732}

\bibitem[\protect\citeauthoryear{{Reboul-Salze}, {Guilet}, {Raynaud}  \&
  {Bugli}}{{Reboul-Salze} et~al.}{2020}]{Reboul20}
{Reboul-Salze} A.,  {Guilet} J.,  {Raynaud} R.,   {Bugli} M.,  2020, arXiv
  e-prints, \href {https://ui.adsabs.harvard.edu/abs/2020arXiv200503567R} {p.
  arXiv:2005.03567}

\bibitem[\protect\citeauthoryear{{Reboul-Salze}, {Guilet}, {Raynaud}  \&
  {Bugli}}{{Reboul-Salze} et~al.}{2022}]{Reboul22}
{Reboul-Salze} A.,  {Guilet} J.,  {Raynaud} R.,   {Bugli} M.,  2022, \mn@doi
  [\aap] {10.1051/0004-6361/202142368}, \href
  {https://ui.adsabs.harvard.edu/abs/2022A&A...667A..94R} {667, A94}

\bibitem[\protect\citeauthoryear{{Rheinhardt} \& {Geppert}}{{Rheinhardt} \&
  {Geppert}}{2005}]{Rheinhardt05}
{Rheinhardt} M.,  {Geppert} U.,  2005, \mn@doi [\aap]
  {10.1051/0004-6361:20042062}, \href
  {https://ui.adsabs.harvard.edu/abs/2005A&A...435..201R} {435, 201}

\bibitem[\protect\citeauthoryear{{Sawai} \& {Yamada}}{{Sawai} \&
  {Yamada}}{2016}]{Sawai16}
{Sawai} H.,  {Yamada} S.,  2016, \mn@doi [\apj] {10.3847/0004-637X/817/2/153},
  \href {https://ui.adsabs.harvard.edu/abs/2016ApJ...817..153S} {817, 153}

\bibitem[\protect\citeauthoryear{{Smartt}}{{Smartt}}{2009}]{Smartt09}
{Smartt} S.~J.,  2009, \mn@doi [\araa] {10.1146/annurev-astro-082708-101737},
  \href {https://ui.adsabs.harvard.edu/abs/2009ARA&A..47...63S} {47, 63}

\bibitem[\protect\citeauthoryear{{Spruit}, {Nordlund}  \& {Title}}{{Spruit}
  et~al.}{1990}]{Spruit90}
{Spruit} H.~C.,  {Nordlund} A.,   {Title} A.~M.,  1990, \mn@doi [\araa]
  {10.1146/annurev.aa.28.090190.001403}, \href
  {https://ui.adsabs.harvard.edu/abs/1990ARA&A..28..263S} {28, 263}

\bibitem[\protect\citeauthoryear{{Sugiura}, {Furusawa}, {Sumiyoshi}  \&
  {Yamada}}{{Sugiura} et~al.}{2022}]{Sugiura2022}
{Sugiura} K.,  {Furusawa} S.,  {Sumiyoshi} K.,   {Yamada} S.,  2022, \mn@doi
  [Progress of Theoretical and Experimental Physics] {10.1093/ptep/ptac118},
  \href {https://ui.adsabs.harvard.edu/abs/2022PTEP.2022k3E01S} {2022, 113E01}

\bibitem[\protect\citeauthoryear{{Summa}, {Hanke}, {Janka}, {Melson}, {Marek}
  \& {M{\"u}ller}}{{Summa} et~al.}{2016}]{Summa16}
{Summa} A.,  {Hanke} F.,  {Janka} H.-T.,  {Melson} T.,  {Marek} A.,
  {M{\"u}ller} B.,  2016, \mn@doi [\apj] {10.3847/0004-637X/825/1/6}, \href
  {https://ui.adsabs.harvard.edu/abs/2016ApJ...825....6S} {825, 6}

\bibitem[\protect\citeauthoryear{{Suwa}, {Kotake}, {Takiwaki}, {Whitehouse},
  {Liebend{\~A}-rfer}  \& {Sato}}{{Suwa} et~al.}{2010}]{Suwa10}
{Suwa} Y.,  {Kotake} K.,  {Takiwaki} T.,  {Whitehouse} S.~C.,
  {Liebend{\~A}-rfer} M.,   {Sato} K.,  2010, \mn@doi [\pasj]
  {10.1093/pasj/62.6.L49}, \href
  {https://ui.adsabs.harvard.edu/abs/2010PASJ...62L..49S} {62, L49}

\bibitem[\protect\citeauthoryear{{Taddia} et~al.,}{{Taddia}
  et~al.}{2018}]{Taddia18}
{Taddia} F.,  et~al., 2018, \mn@doi [\aap] {10.1051/0004-6361/201730844}, \href
  {https://ui.adsabs.harvard.edu/abs/2018A&A...609A.136T} {609, A136}

\bibitem[\protect\citeauthoryear{{Takahashi} \& {Langer}}{{Takahashi} \&
  {Langer}}{2021}]{Takahashi21}
{Takahashi} K.,  {Langer} N.,  2021, \mn@doi [\aap]
  {10.1051/0004-6361/202039253}, \href
  {https://ui.adsabs.harvard.edu/abs/2021A&A...646A..19T} {646, A19}

\bibitem[\protect\citeauthoryear{{Takiwaki}, {Kotake}, {Nagataki}  \&
  {Sato}}{{Takiwaki} et~al.}{2004}]{Takiwaki04}
{Takiwaki} T.,  {Kotake} K.,  {Nagataki} S.,   {Sato} K.,  2004, \mn@doi [\apj]
  {10.1086/424993}, \href
  {https://ui.adsabs.harvard.edu/abs/2004ApJ...616.1086T} {616, 1086}

\bibitem[\protect\citeauthoryear{{Takiwaki}, {Kotake}  \& {Sato}}{{Takiwaki}
  et~al.}{2009}]{Takiwaki09}
{Takiwaki} T.,  {Kotake} K.,   {Sato} K.,  2009, \mn@doi [\apj]
  {10.1088/0004-637X/691/2/1360}, \href
  {https://ui.adsabs.harvard.edu/abs/2009ApJ...691.1360T} {691, 1360}

\bibitem[\protect\citeauthoryear{{Takiwaki}, {Kotake}  \& {Suwa}}{{Takiwaki}
  et~al.}{2012}]{Takiwaki12}
{Takiwaki} T.,  {Kotake} K.,   {Suwa} Y.,  2012, \mn@doi [\apj]
  {10.1088/0004-637X/749/2/98}, \href
  {https://ui.adsabs.harvard.edu/abs/2012ApJ...749...98T} {749, 98}

\bibitem[\protect\citeauthoryear{{Takiwaki}, {Kotake}  \& {Suwa}}{{Takiwaki}
  et~al.}{2016}]{Takiwaki16}
{Takiwaki} T.,  {Kotake} K.,   {Suwa} Y.,  2016, \mn@doi [\mnras]
  {10.1093/mnrasl/slw105}, \href
  {https://ui.adsabs.harvard.edu/abs/2016MNRAS.461L.112T} {461, L112}

\bibitem[\protect\citeauthoryear{{Thompson} \& {Duncan}}{{Thompson} \&
  {Duncan}}{1993}]{Thompson93}
{Thompson} C.,  {Duncan} R.~C.,  1993, \mn@doi [\apj] {10.1086/172580}, \href
  {https://ui.adsabs.harvard.edu/abs/1993ApJ...408..194T} {408, 194}

\bibitem[\protect\citeauthoryear{{Varma} \& {M{\"u}ller}}{{Varma} \&
  {M{\"u}ller}}{2021}]{Varma21}
{Varma} V.,  {M{\"u}ller} B.,  2021, \mn@doi [\mnras] {10.1093/mnras/stab883},
  \href {https://ui.adsabs.harvard.edu/abs/2021MNRAS.504..636V} {504, 636}

\bibitem[\protect\citeauthoryear{{Varma}, {M{\"u}ller}  \& {Schneider}}{{Varma}
  et~al.}{2023}]{Varma23}
{Varma} V.,  {M{\"u}ller} B.,   {Schneider} F. R.~N.,  2023, \mn@doi [\mnras]
  {10.1093/mnras/stac3247}, \href
  {https://ui.adsabs.harvard.edu/abs/2023MNRAS.518.3622V} {518, 3622}

\bibitem[\protect\citeauthoryear{{Vartanyan}, {Coleman}  \&
  {Burrows}}{{Vartanyan} et~al.}{2022}]{Vartanyan22}
{Vartanyan} D.,  {Coleman} M. S.~B.,   {Burrows} A.,  2022, \mn@doi [\mnras]
  {10.1093/mnras/stab3702}, \href
  {https://ui.adsabs.harvard.edu/abs/2022MNRAS.510.4689V} {510, 4689}

\bibitem[\protect\citeauthoryear{{White}, {Burrows}, {Coleman}  \&
  {Vartanyan}}{{White} et~al.}{2022}]{White22}
{White} C.~J.,  {Burrows} A.,  {Coleman} M. S.~B.,   {Vartanyan} D.,  2022,
  \mn@doi [\apj] {10.3847/1538-4357/ac4507}, \href
  {https://ui.adsabs.harvard.edu/abs/2022ApJ...926..111W} {926, 111}

\bibitem[\protect\citeauthoryear{{Woosley}, {Heger}  \& {Weaver}}{{Woosley}
  et~al.}{2002}]{Woosley02}
{Woosley} S.~E.,  {Heger} A.,   {Weaver} T.~A.,  2002, \mn@doi [Reviews of
  Modern Physics] {10.1103/RevModPhys.74.1015}, \href
  {https://ui.adsabs.harvard.edu/abs/2002RvMP...74.1015W} {74, 1015}

\bibitem[\protect\citeauthoryear{{Yoshida}, {Takiwaki}, {Kotake}, {Takahashi},
  {Nakamura}  \& {Umeda}}{{Yoshida} et~al.}{2021}]{Yoshida21}
{Yoshida} T.,  {Takiwaki} T.,  {Kotake} K.,  {Takahashi} K.,  {Nakamura} K.,
  {Umeda} H.,  2021, \mn@doi [\apj] {10.3847/1538-4357/abd3a3}, \href
  {https://ui.adsabs.harvard.edu/abs/2021ApJ...908...44Y} {908, 44}

\makeatother
\end{thebibliography}

\label{lastpage}
\end{document}